\documentclass[12pt]{article}

\usepackage{graphicx}
\usepackage{latexsym}

\title{ Hypermultiplet dependence of the effective
action in  ${\cal N}=2$ superconformal theories }

\author{N.G. Pletnev\footnote{pletnev@math.nsc.ru} }
\date{}

\date{\it
Department of
Theoretical Physics,\\ Institute of Mathematics, Novosibirsk , \\
630090, Russia\\}

\begin{document}

\maketitle

\begin{abstract}
I review the approach \cite{jhep} to the one-loop low-energy
effective action in the hypermultiplet sector for ${\cal N}=2$
superconformal models. Any such a model contains an ${\cal N}=2$
vector multiplet and some number of hypermultiplets. We found a
general expression for the low-energy effective action in the form
of a proper-time integral. The leading space-time dependent
contributions to the effective action are derived and their
bosonic component structure is analyzed. The component action
contains terms with three and four space-time derivatives of
component fields and has the Chern-Simons-like form.

\end{abstract}

\thispagestyle{empty}
\newcommand{\be}{\begin{equation}}
\newcommand{\ee}{\end{equation}}
\newcommand{\bea}{\begin{eqnarray}}
\newcommand{\eea}{\end{eqnarray}}

\section{Introduction}
I am very glad to take part in this book devoted to celebration of
the 60 birth day of remarkable scientist and my dear friend Ioseph
L. Buchbinder.

Four-dimensional ${\cal N}=2$ supersymmetric gauge theories are
formulated in terms of ${\cal N}=2$ vector multiplet coupled to a
massless hypermultiplets in certain representations ${ R}$ of the
gauge group $G$. All such models possess only one-loop divergences
\cite{howe} and can be made finite at certain restrictions on
representations and field contents. In the model with $n_\sigma$
hypermultiplets in representations $R_\sigma$ of the gauge group
${G}$ the finiteness condition has simple and universal form
\begin{equation}\label{fin}
  C({G})=\sum_\sigma n_\sigma T({R_\sigma}),
\end{equation}
where $C(G)$ is the quadratic Casimir operator for the adjoint
representation and $T(R_\sigma)$ is the quadratic Casimir operator
for the representation ${R_\sigma}$. A simplest solution to
Eq.(\ref{fin}) is ${\cal N}=4$ SYM theory where $n_\sigma=1$ and
all fields are taken in the adjoint representation. It is evident
that there are other solutions, e.g. for the case of $\mathop{\rm
SU}(N)$ group and hypermultiplets in the fundamental
representation one gets $T(R)=1/2$, $C(G) =N$ and $n_\sigma = 2N$.
A number of ${\cal N}=2$ superconformal models has been
constructed in the context of AdS/CFT correspondence (see e.g.
\cite{ADS}, the examples of such models and description of
structure of vacuum states were discussed in details e.g. in Ref.
\cite{KMT} ).

In this paper we study the structure of the low-energy one-loop
effective action for the ${\cal N}=2$ superconformal theories. The
effective action of the ${\cal N}=4$ SYM theory and ${\cal N}=2$
superconformal models in the sector of ${\cal N}=2$ vector
multiplet has been studied by various methods. However a problem
of hypermultiplet dependence of the effective action in the above
theories was open for a long time.

The low-energy effective action containing both ${\cal N}=2$
vector multiplet and hypermultiplet background fields in ${\cal
N}=4$ SYM theory was first constructed in Ref. \cite{BI} and
studied in more details in \cite{BP}. In this paper we will
consider the hypermultiplet dependence of the effective action for
${\cal N}=2$ superconformal models. Such models are finite
theories as well as the ${\cal N}=4$ SYM theory and one can expect
that hypermultiplet dependence of the effective action in ${\cal
N}=2$ superconformal models is analogous to one in ${\cal N}=4$
SYM theory. However this is not so evident. The ${\cal N}=4$ SYM
theory is a special case of the ${\cal N}=2$ superconformal
models, however it possesses extra ${\cal N}=2$ supersymmetry in
comparison with generic ${\cal N}=2$ models. As it was noted in
\cite{BI} just this extra ${\cal N}=2$ supersymmetry is the key
point for finding an explicit hypermultiplet dependence of the
effective action in ${\cal N}=4$ SYM theory. Therefore a
derivation of the effective action for ${\cal N}=2$ superconformal
models in the hypermultiplet sector is an independent problem.

In this paper we derive the complete ${\cal N}=2$ supersymmetric
one-loop effective action depending both on the background vector
multiplet and hypermultiplet fields in a mixed phase where both
vector multiplet and hypermultiplet have non-vanishing expectation
values. The ${\cal N}=2$ supersymmetric models under consideration
are formulated in  harmonic superspace \cite{gios}. We develop a
systematic method of constructing the lower- and higher-derivative
terms in the one-loop effective action given in terms of a heat
kernel for certain differential operators on the harmonic
superspace and calculate the heat kernel depending on ${\cal N}=2$
vector multiplet and hypermultiplet background superfields. We
study a component form of a leading quantum corrections for
on-shell and beyond on-shell background hypermultiplets and find
that they contain, among the others, the terms corresponding to
the Chern-Simons-type actions. The necessity of such manifest
scale invariant $P$-odd terms in effective action of ${\cal N}=4$
SYM theory, involving both scalars and vectors, has been pointed
out in \cite{TZ}. Proposal for the higher-derivative terms in the
effective action of the ${\cal N}=2$ models in the harmonic
superspace has been given in \cite{arg}. We show how the terms in
the effective action assumed in P.C. Argyres at al. can be
actually computed in supersymmetric quantum field theory.

\section{The model and background field splitting}

${\cal N}=2$ harmonic superspace has been introduced in
\cite{gikos} extending the standard ${\cal N}=2$ superspace with
coordinates ${ z^M=(x^m,\theta^\alpha_i,
\bar\theta^i_{\dot\alpha})}$ ($i =1,2$) by the harmonics ${
u^{\pm}_{i}}$ parameterizing the two-dimensional sphere $S^2$: $
u^{+i}u^-_i=1, \quad \overline{u^{+i}}=u^-_i.$

The main advantage of harmonic superspace is that the ${\cal N}=2$
vector multiplet and hypermultiplet can be described by
unconstrained superfields over the analytic subspace with the
coordinates $ \zeta^M \equiv (x^m_A,
\theta^{+\alpha},\bar\theta^+_{\dot\alpha}, u^{\pm}_i),$ where the
so-called analytic basis is defined by
\begin{equation}\label{analyt}
x^m_A=x^m-i\theta^+\sigma\bar\theta^-
-i\theta^-\sigma^m\bar\theta^+, \quad
\theta^{\pm}_\alpha=u^{\pm}_i\theta^i_\alpha,\quad
\bar\theta^{\pm}_{\dot\alpha}=u^{\pm}_i\bar\theta^i_{\dot\alpha}~.
\end{equation}
The ${\cal N}=2$ vector multiplet is described by a real  analytic
superfield ${ V^{++}=V^{++I}(\zeta)T_I}$ taking values in the Lie
algebra of the gauge group. A hypermultiplet, transforming in the
representation $R$ of the gauge group, is described by an analytic
superfield ${\bf q^+(\zeta)}$ and its conjugate ${\bf
\tilde{q}^+(\zeta)}$ .

The classical action of ${\cal N}=2$ SYM theory coupled to
hypermultiplets consist of two parts: the pure ${\cal N}=2$ SYM
action  and the $q$-hypermultiplet action in the fundamental or
adjoint representation of the gauge group. Written in the harmonic
superspace  its action  reads
\begin{equation}\label{class}
 S=\frac{1}{2g^2}\mbox{tr}\int d^8z \,{\cal W}^2
+\frac{1}{2}\int d \zeta^{(-4)} q^{+f}_{a}(D^{++}
+igV^{++})q^{+a}_{f}~,
\end{equation}
where we used the doublet notation $q^+_a=(q^+, -\tilde{q}^+)$. By
construction, the action (\ref{class}) is manifestly ${\cal N}=2$
supersymmetric. Here $d\zeta^{(-4)}=d^4x d^4\theta^+ du$ denotes
the analytic subspace integration measure and $$ {\cal
D}^{++}=D^{++}+iV^{++}, \quad
D^{++}=\partial^{++}-2i\theta^+\sigma^m\bar\theta^+\partial_m,
\quad
\partial^{++}\equiv u^{+i}\frac{\partial}{\partial u^{-i}}$$ is the
analyticity-preserving covariant harmonic derivative. It can be
shown that $V^{++}$ is the single unconstrained analytic,
$D^+_{(\alpha,\dot\alpha)}V^{++}=0$, prepotential of the pure
${\cal N}=2$ SYM theory, and all other geometrical object are
determined in terms of it. So,the covariantly chiral superfield
strength ${\cal W}$
\begin{equation}\label{str}
 {\cal W}=-\frac{1}{4}(\bar{D}^+)^2 V^{--}, \quad \bar{\cal
W}=-\frac{1}{4}(D^+)^2 V^{--}.
\end{equation}
is expressed through the (nonanalytic) real superfield $V^{--}$
satisfying the equation
$$D^{++}V^{--}-D^{--}V^{++}+i[V^{++}, V^{--}]=0.$$ This equation
has a solution in form of the power series in $V^{++}$
\cite{zupnik}.

For further use we will write down also the superalgebra of gauge
covariant derivatives with the notation ${\cal
D}^{\pm}_{(\alpha,\dot\alpha)}={\cal
D}^{i}_{(\alpha,\dot\alpha)}u^{\pm}_i$:
\begin{equation}\label{alg}
\{{\cal D}^+_\alpha, {\cal
D}^-_\beta\}=-2i\varepsilon_{\alpha\beta}\bar{\cal W}~, \quad
\{\bar{\cal D}^+_{\dot\alpha}, \bar{\cal
D}^-_{\dot\beta}\}=2i\varepsilon_{\dot\alpha\dot\beta}{\cal W}~,
\end{equation}
$$
\{\bar{\cal D}^+_{\dot\alpha}, {\cal D}^-_\alpha\}=-\{{\cal
D}^+_{\alpha}, \bar{\cal D}^-_{\dot\alpha}\}=2i{\cal
D}_{\alpha\dot\alpha}~,
$$
$$
[{\cal D}^{\pm}_\alpha, {\cal
D}_{\beta\dot\beta}]=\varepsilon_{\alpha\beta}\bar{\cal
D}^{\pm}_{\dot\beta}\bar{\cal W}~, \quad [\bar{\cal
D}^{\pm}_{\dot\alpha}, {\cal
D}_{\beta\dot\beta}]=\varepsilon_{\dot\alpha\dot\beta}{\cal
D}^{\pm}_{\beta}{\cal W}~,
$$
$$
[{\cal D}_{\alpha\dot\alpha}, {\cal
D}_{\beta\dot\beta}]=\frac{1}{2i}\{\varepsilon_{\alpha\beta}\bar{\cal
D}^+_{\dot\alpha}\bar{\cal D}^-_{\dot\beta}\bar{\cal W} +
\varepsilon_{\dot\alpha\dot\beta}{\cal D}^-_{\alpha}{\cal
D}^+_{\beta}{\cal
W}\}=\frac{1}{2i}\{\varepsilon_{\alpha\beta}\bar{F}_{\dot\alpha\dot\beta}
+ \varepsilon_{\dot\alpha\dot\beta}F_{\alpha\beta}\}~.
$$
The operators ${\cal D}^+_\alpha$ and $\bar{\cal
D}^+_{\dot\alpha}$ strictly anticommute
\begin{equation}\label{d+}
\{{\cal D}^+_\alpha,{\cal D}^+_\beta\}=\{\bar{\cal
D}^+_{\dot\alpha},\bar{\cal D}^+_{\dot\beta}\}=\{{\cal
D}^+_\alpha,\bar{\cal D}^+_{\dot\alpha}\}=0~.
\end{equation}
A full set of gauge covariant derivatives includes also the
harmonic derivatives $({\cal D}^{++},\\ {\cal D}^{--}, {\cal
D}^{0})$, which form the algebra $su(2)$ and satisfy the obviously
commutation relations with ${\cal D}^{\pm}_\alpha$ and $\bar{\cal
D}^{\pm}_{\dot\alpha}$.

The action (\ref{class}) possesses the superconformal symmetry
$SU(2,2|2)$ which is manifest in the harmonic superspace approach.
The low energy effective action at a generic vacuum of ${\cal
N}=2$ gauge theory includes only massless $\mathop{\rm U}(1)$
vector multiplets and massless neutral hypermultiplets, since
charged vectors and charged hypermultiplets get masses by the
Higgs mechanism. The moduli space of vacua for the theory under
consideration is specified by the following conditions
\cite{modul}:
\begin{equation}\label{vacua}
{ [\bar\phi, \phi]=0, \quad \phi f_i=0, \quad \bar{f}^i \bar\phi
=0 \quad \bar{f}^{(i} T_I f^{j)}=0}~.
\end{equation}
Here the $\phi, \bar\phi$ are the scalar components of ${\cal
N}=2$ vector multiplet and complex scalars $f_{i}$  are the scalar
components of the hypermultiplet.

The structure of a vacuum state is characterized by solutions to
Eqs. (\ref{vacua}). These solutions can be classified according to
the phases or branches of the gauge theory under consideration. In
the pure Coulomb phase  ${ f_i =0}$, ${ \phi \not= 0}$ and
unbroken gauge group is $\mathop{\rm U}(1)^{ \rm rank({ G})}$. In
the pure Higgs phase ${ f_i \not=0}$ and the gauge symmetry is
completely broken; there are no massless gauge bosons.  In the
mixed phases, i.e. on the direct product of the Coulomb and Higgs
branches (some number of $\phi, \bar\phi$ is not equal to zero and
some number of $f_{i}$ is not equal to zero) the gauge group is
broken down to ${\tilde{G}} \times K$ where $K$ is some Abelian
subgroup.

Further we  impose the special restrictions on the background
${\cal N}=2$ vector multiplet and hypermultiplet. They are chosen
to be aligned along a fixed direction in the moduli space vacua;
in particular, their scalar fields should solve Eqs.
(\ref{vacua}):
\begin{equation}\label{vac}
V^{++}={\bf V}^{++}(\zeta){H}, \quad q^+={\bf q}^+
(\zeta)\Upsilon~.
\end{equation}
Here ${H}$ is a fixed generator in the Cartan subalgebra
corresponding to Abelian subgroup $K$, and $\Upsilon$ is a fixed
vector in the ${ R}$-representation space of the gauge group,
where the hypermultiplet takes values, chosen so that ${H}
\Upsilon=0$ and $\bar\Upsilon { T}_I \Upsilon =0.$ Eq.(\ref{vac})
defines a single $\mathop{\rm U}(1)$ vector multiplet and a single
hypermultiplet which is neutral with respect to the $\mathop{\rm
U}(1)$ gauge subgroup generated by ${ H}$.

At the tree level and energies below the symmetry breaking scale,
we have free field massless dynamics of the ${\cal N}=2$ vector
multiplet and the hypermultiplet aligned in a particular direction
in the moduli space of vacua. Thus the low energy propagating
fields are massless neutral hypermultiplets and $\mathop{\rm
U}(1)$ vector which form the on shell superfields possessing the
properties
\begin{equation}\label{onsh}
(D^{\pm})^2{\cal W}=(\bar{D}^{\pm})^2\bar{\cal W}=0~,
\end{equation}
$$
D^{++}q^{+a}=(D^{--})^2q^{+a}=D^{--}q^{-a}=0, \quad
q^{-a}=D^{--}q^{+a}, \quad D^-_{(\alpha,\dot\alpha)}q^{-a}=0~.
$$
The equations (\ref{onsh}) eliminate the auxiliary fields and put
the physical fields on shell.

At the quantum level, however, exchanges of virtual massive
particles produce the corrections to the action of the massless
fields.  We quantize the ${\cal N}=2$ supergauge theory in the
framework of the ${\cal N}=2$ supersymmetric background field
method \cite{echaya}  by splitting the fields $V^{++}, q^{+a}$
into the sum of the background fields $V^{++}, q^{+a}$,
parameterized according to (\ref{vac}), and the quantum fields
$v^{++}, Q^{+a}$ and expanding the Lagrangian in a power series in
quantum fields. Such a procedure allows us to find the effective
action for arbitrary ${\cal N}=2$ supersymmetric gauge model in a
form preserving the manifest ${\cal N}=2$ supersymmetry and
classical gauge invariance in quantum theory.

In the background-quantum splitting, the classical action of the
pure ${\cal N}=2$ SYM theory can be shown to be given by
\begin{equation}\label{splitt}
S_{SYM}[V^{++}+v^{++}]=S_{SYM}[V^{++}]+\frac{1}{4}\int
d\zeta^{(-4)}duv^{++}(D^+)^2{\cal W}_\lambda
\end{equation}
$$
-\mbox{tr}\int d^{12}z \sum_{n=2}^\infty \frac{(-ig)^{n-2}}{n}\int
du_1...du_n\frac{v^{++}_\tau(z,u_1)...v^{++}_\tau(z,u_n)}{(u_1^+u_2^+)...(u_n^+u_1^+)}~.
$$
${\cal W}_\lambda$ and $v^{++}_\tau$ denote the $\lambda$- and
$\tau$-frame forms of $\cal W$ and $v^{++}$ respectively.   The
hypermultiplet action becomes
\begin{equation}\label{hypeq}
S_H(q+Q)=S_H[q] +\int d\zeta^{(-4)}du Q^+_a{\cal
D}^{++}q^{+a}+\frac{1}{2}\int d\zeta^{(-4)}duq^+_a iv^{++}q^{+a}
\end{equation}
$$
+\frac{1}{2}\int d\zeta^{(-4)}du\{Q^+_a{\cal D}^{++}Q^{+a}+Q^+_a
iv^{++}q^{+a}+q^+_a iv^{++}Q^{+a}+Q^+_a iv^{++}Q^{+a}\}~.
$$
The terms linear in $v^{++}$ and $q^+$ in (\ref{splitt}),
(\ref{hypeq}) determines the equation of motion and this term
should be dropped when considering the effective action.

To construct the effective action, we will follow the
Faddeev-Popov Ansatz.  We  write the final result for the
effective action $\Gamma[V^{++}, q^+]$
\begin{equation}\label{path3}
e^{i\Gamma[V^{++},\; q^+]}=e^{iS_{cl}[V^{++},\;
q^+]}\mbox{Det}^{1/2}{\stackrel{\frown}{\Box}}_{(4,0)} \int {\cal
D}v^{++}{\cal D}Q^+ {\cal D}{\bf b}{\cal D}{\bf c}{\cal D}\varphi
e^{iS_{q}},\end{equation} where
${\stackrel{\frown}{\Box}}=-\frac{1}{2}({\cal D}^+)^4({\cal
D}^{--})^2 $ and action $S_{q}$ is as follows
$$
S_{q}[v^{++}, Q^+, {\bf b}, {\bf c}, \varphi, V^{++},
q^+]=S_{2}[v^{++}, Q^+, {\bf b}, {\bf c}, \varphi, V^{++},
q^+]+S_{int},
$$
\begin{equation}\label{q2}
S_2=-\frac{1}{2}\mbox{tr}\int d\zeta^{(-4)}du
v^{++}{\stackrel{\frown}{\Box}} v^{++} +\mbox{tr}\int
d\zeta^{(-4)}du{\bf b}({\cal D}^{++})^2{\bf c}\end{equation}
$$+\frac{1}{2}\mbox{tr}\int d\zeta^{(-4)}du\varphi({\cal
D}^{++})^2\varphi +\frac{1}{2}\int d\zeta^{(-4)}du\{Q^+_a{\cal
D}^{++}Q^{+a}$$$$+Q^+_a iv^{++}q^{+a}+q^+_a iv^{++}Q^{+a}\}~,
$$
This equations completely determine the structure of the
perturbation expansion for calculating the effective action
$\Gamma[V^{++}, q^+]$ of the ${\cal N}=2$ SYM theory with
hypermultiplets in a manifestly supersymmetric and gauge invariant
form. The action $S_2$ defines the propagators depending on
background fields. In the framework of the background field
formalism in ${\cal N}=2$ harmonic superspace there appear three
types of covariant matter and gauge field propagators. Associated
with $\stackrel{\frown}{\Box}$ is a Green's function
$G^{(2,2)}(z,z')$ which  satisfies the equation
$\stackrel{\frown}{\Box}G^{(2,2)}(1|2)=-{\bf
1}\delta^{(2,2)}(1|2)$, is
\begin{equation}\label{provec}
G^{(2,2)}(1,2)=-\frac{1}{2\stackrel{\frown}{\Box}_1\stackrel{\frown}{\Box}_2}({\cal
D}^+_1)^4({\cal D}^+_2)^4\{{\bf
1}\delta^{12}(z_1-z_2)(D^{--}_2)^2\delta^{(-2,2)}(u_1,u_2)\}~.
\end{equation}
The $Q^+$ hypermultiplet propagator associated with the action
(\ref{q2}) has the form
\begin{equation}\label{prohyp}
 G_b^{a (1.1)}(1|
2)=-\delta^a_b \frac{({\cal D}^+_1)^4({\cal
D}^+_2)^4}{(u^+_1u^+_2)^3}\frac{1}{{\stackrel{\frown}{\Box}}_1}
\delta^{12}(z_1-z_2)~.
\end{equation}
It is not hard to see that this manifestly analytic expression is
the solution of the equation ${\cal
D}^{++}_1G^{(1,1)}=\delta_A^{(3,1)}(1|2).$ For the hypermultiplet
of the second type described by a chargeless real analytic
superfield $\omega(\zeta,u)$ the equation for Green' function is\\
$({\cal D}^{++}_1)^2G^{(0,0)}(1|2)=\delta_A^{(4,0)}(1|2)$. The
suitable expression for $G^{(0,0)}$ is
\begin{equation}\label{proomeg}
G^{(0,0)}(1|2)=-\frac{1}{{\stackrel{\frown}{\Box}}_1}({\cal
D}^+_1)^4({\cal D}^+_2)^4\{{\bf
1}\delta^{12}(z_1-z_2)\frac{u^-_1u^-_2}{(u^+_1u^+_2)^3}\}.
\end{equation}
The operator ${\stackrel{\frown}{\Box}}=-\frac{1}{2}({\cal
D}^+)^4({\cal D}^{--})^2$ transforms each covariantly analytic
superfield into a covariantly analytic and, using algebra
(\ref{alg}), can be rewritten as second-order d'Alemberian-like
differential operator on the space of such superfields. The
coefficients of this operator depend on background super\-fields
${\cal W}, \bar{\cal W}$.

\section{Structure of the one-loop effective action}
Consider the loop expansion of the effective action within the
background field formulation.  A formal expression of the one-loop
effective action $\Gamma[V^{++}, \; q^+]$ for the theory under
consideration is written in terms of a path integral as follows
(\ref{path3}), where the full quadratic action is defined in Eq.
(\ref{q2}). Here $v^{++}$ is a quantum vector superfield taking
values in the Lie algebra of the gauge group and ${\bf b}$, ${\bf
c}$ are two real analytic Faddeev-Popov fermionic ghosts and
$\varphi$ is the bosonic Nielsen-Kallosh ghost, all in the adjoint
representation of the gauge group.

In the vector sector of the ${\cal N}=2$ SYM theory where the
matter hypermultiplet are integrated out, the one-loop effective
action $\Gamma [V^{++}]$ reads
\begin{equation}
\Gamma
[V^{++}]=\frac{i}{2}\mbox{Tr}_{(2,2)}\ln\stackrel{\frown}{\Box}
-\frac{i}{2}\mbox{Tr}_{(4,0)}\ln\stackrel{\frown}{\Box}-\frac{i}{2}\mbox{Tr}_{ad}\ln({\cal
D}^{++})^2 +i\mbox{Tr}_{R_q}\ln{\cal D}^{++}
+\frac{i}{2}\mbox{Tr}_{R_\omega}\ln({\cal D}^{++})^2.
\end{equation}
Currently, the holomorphic and non-holomorphic
parts of the low-energy effective action ${\cal N}=2,4$ SYM theory
on the Coulomb branch, including Heisenberg-Euler type action in
the presence of a covariantly constant vector multiplet, are
completely known. The general structure of the low-energy
effective action in ${\cal N}=2,4$ superconformal theories is
\cite{BKT}:
$$
\Gamma=S_{cl}+c\int d^{12}z\ln{\cal W}\ln\bar{\cal W}+\int
d^{12}z\ln{\cal W}\Lambda(\frac{D^4\ln{\cal W}}{\bar{\cal
W}^2})+c.c. +\int d^{12}z\Upsilon(\frac{\bar{D}^4\ln\bar{\cal
W}}{{\cal W}^2},\frac{D^4\ln{\cal W}}{\bar{\cal W}^2}),
$$
where $\Lambda$ and $\Upsilon$ are holomorphic and real analytic
function of the (anti)chiral superconformal invariants. The
$c$-term is known to generate four-derivative quantum corrections
at the component level which include an famous $F^4$ term.

The hypermultiplet dependent part of the effective action in
${\cal N}=4$ SYM theory in leading order is also known \cite{BIP}.
For further analysis of the effective action it is convenient to
diagonalize the action of quantum fields $S^{(2)}$ using a special
shift of hypermultiplet variables in the path integral
\begin{equation}\label{replac}
Q^{+a}= \xi^{+a} + i\int d \zeta^{(-4)}_2  q^{+b}(2)v^{++}(2)
G_b^{a (1.1)}(1|2),\end{equation}$$ Q^{+}_a= \xi^{+}_a - i\int d
\zeta^{(-4)}_2 G_a^{b (1.1)}(1|2) v^{++}(2)q^{+}_{b}(2)~,$$ where
$\xi^{+a}, \xi^{+}_{a}$ are the new independent variables in the
path integral. It is evident that the Jacobian of the replacement
(\ref{replac}) is equal to unity. Here $G_b^{a (1.1)}(1| 2)$ is
the background-dependent propagator (\ref{prohyp}) for the
superfields $Q^{+a}, Q^+_b$. In terms of the new set of quantum
fields we obtain for the following hypermultiplet dependent part
of the quadratic action
\begin{equation}
S^{(2)}_H=-\frac{1}{2}\int d\zeta^{(-4)} \xi^{a+}{\cal
D}^{++}\xi^+_a -\frac{1}{2}\int d\zeta^{(-4)}_1 d\zeta^{(-4)}_2
q^{+a}(1)v^{++}(1) G_a^{b (1.1)}(1|2)v^{++}(2)q^+_b(2)~.
\end{equation}
Then the vector multiplet dependent part of the
quadratic action gets the following non-local extension
\begin{equation}\label{green2}
S^{(2)}_v=-\frac{1}{2}\mbox{tr}\int d \zeta^{(-4)}_1 v^{++}_1 \int
d \zeta^{(-4)}_2\left(
{\stackrel{\frown}{\Box}}\delta^{(2.2)}_A(1|2)
+q^{+a}(1)G_a^{b(1.1)}(1|2)q^+_b(2) \right) v^{++}_2~.
\end{equation}
Expression (\ref{green2}), written as an analytical nonlocal
superfunctional, will be a starting point for our calculations of
the one-loop effective action in the hypermultiplet sector. Our
aim in the current and later sections is to find the leading
low-energy contribution to the effective action for the slowly
varying hypermultiplet when all derivatives of the background
hypermultiplet can be neglected. We will show that for such a case
the non-local interaction is localized.

Using the relation $v^{++}_2=\int d\zeta_3^{(-4)} \delta^{(2.2)}_A
(2|3)v^{++}_3$ one can rewrite expression for $S^{(2)}_v$
(\ref{green2}) in the form
\begin{equation}\label{nonlocal}
S^{(2)}_v=-\frac{1}{2}\mbox{tr}\int d \zeta^{(-4)}_1\, v^{++}_1
\int d \zeta^{(-4)}_2(
{\stackrel{\frown}{\Box}}\delta^{(2.2)}_A(1|2)\end{equation}$$
+\int d\zeta^{(-4)}_3
q^{+a}(1)G_a^{b(1.1)}(1|3)q^+_b(3)\delta^{(2.2)}_A(3|2)
v^{++}_2)~.
$$
Then we use the explicit form of the Green function (\ref{prohyp})
and the relation allowing us to express the $({\cal
D}^{+}_1)^4({\cal D}^+_2)^4$ as a polynomial in powers of
$(u^+_1u^+_2)$ \cite{KMcA}:
\begin{equation}\label{polin}
 ({\cal D}^+_1)^4({\cal D}^+_2)^4\end{equation}$$=
({\cal D}^+_1)^4\left(({\cal D}^-_1)^4(u^+_1u^+_2)^4
-\frac{i}{2}\Delta_1^{--}(u^+_1u^+_2)^3(u^-_1u^+_2)
-{\stackrel{\frown}{\Box}}_1(u^+_1u^+_2)^2(u^-_1u^+_2)^2\right)~,
$$
where the operator $\Delta^{--}$ is
\begin{equation}\label{Delt}
\Delta^{--}={\cal D}^{\alpha\dot\alpha}{\cal D}^-_\alpha \bar{\cal
D}^-_{\dot\alpha} +\frac{1}{2}{\cal W}({\cal D}^-)^2
+\frac{1}{2}\bar{\cal W}(\bar{\cal D}^-)^2 +({\cal D}^-{\cal
W}){\cal D}^- +(\bar{\cal D}^-\bar{\cal W})\bar{\cal D}^-~.
\end{equation}
The non-local term in (\ref{nonlocal}) takes the form
$$
\int d\zeta^{(-4)}_3 q^{+a}(1)({\cal
D}^+_3)^4\times$$$$\times\left(({\cal
D}^-_3)^4(u^+_3u^+_1)\frac{1}{{\stackrel{\frown}{\Box}}_3}\right.
\left.-\frac{i}{2}\Delta^{--}_3(u^-_3u^+_1)\frac{1}{{\stackrel{\frown}{\Box}}_3}
-\frac{(u^-_3u^+_1)^2}{(u^+_3u^+_1)}\right)\delta^{12}(1|3)q^+_a(3)\delta^{(2.2)}_A(3|2)~.
$$
The large braces here contain three terms. It is easy to see that
two first terms include the derivatives which will lead to
derivatives of the hypermultiplet in the effective action. Since
we keep only contributions without derivatives, the above terms
can be neglected. As a result, is it sufficient to consider only
the third term in the braces.

Now we apply the relation $\int d\zeta_3^{(-4)}({\cal
D}^+_3)^4=\int d^{12}z,$ allowing to integrate over $z_{3}$, and
obtain
$$-\int du_3\,
q^{+a}(1)\frac{(u_3^-u^+_1)^2}{(u^+_3u^+_1)}q^+_a(u_3, z_1)
\delta^{(2.2)}_A(u_3,z_1|2)~.
$$
Then one uses the on-shell harmonic dependence of hypermultiplet
$q^{+a}(3)=u^+_{3i}q^{ia}$ and take the coincident limit $u_1=u_3$
(conditioned by $\delta^{(2.2)}_A(u_3,z_1|2)$). After that we get
$\int du_3 \frac{u^+_{3 i}}{u^+_3u^+_1}=-u^-_{1 i}$. As a result,
the term under consideration has the form
\begin{equation}q^{+a}(1)q^-_a(1)\delta^{(2.2)}_A(1|2),\end{equation} where the expression
$q^{+a}(1)q^-_a(1)= q^{ia}q_{ia}$ is treated further as the slowly
varying superfield and all its derivatives are neglected. Namely
such an expression was obtained in \cite{BP}  by summation of
harmonic supergraphs.

Thus, the second term in (\ref{nonlocal}) becomes local in the
leading low-energy approximation. As a result, the operator in
action $S^{(2)}_v$ determining the effective background covariant
propagator of the quantum vector multiplet superfield $v^{++}_I$
takes the form
\begin{equation}\label{kin}
\left(\stackrel{\frown}{\Box}_{IJ}
+q^{+a}(z_1,u_1)\{T_I,T_J\}q^-_a(z_1,u_1)\right)\delta^{(2.2)}_A(1|2)~,
\end{equation}
where
\begin{equation}\label{box}
\stackrel{\frown}{\Box}_{IJ} = \mbox{tr}(T_{(I}{\Box}T_{J)}
+\frac{i}{2}T_{(I}[{\cal D}^{+\alpha}{\cal W},T_{J)}]{\cal
D}^-_\alpha
 +\frac{i}{2}T_{(I}[\bar{\cal
D}^+_{\dot\alpha}\bar{\cal W},T_{J)}]\bar{\cal D}^{-\dot\alpha}
+T_{(I}[{\cal W},[\bar{\cal W}, T_{J)}]].
\end{equation}
Here ${\Box} = \frac{1}{2}{\cal D}^{\alpha\dot\alpha}{\cal
D}_{\alpha\dot\alpha}$ is the covariant d'Alemberian.

Thus, using the ${\cal N}=2$ harmonic superspace formulation of
the ${\cal N}=2$ SYM theory with hypermultiplets and techniques of
the non-local shift we obtained that the whole dependence on the
background hypermultiplet is concentrated in the quantum vector
multiplet sector with the modified quadratic action. Therefore the
one-loop effective action is given by the expression
\begin{equation}\label{gamma}
\Gamma^{(1)}[V^{++}, q^+] = \Gamma^{(1)}_v[V^{++}, q^+] +
\widetilde{\Gamma}^{(1)}[V^{++}]~,
\end{equation}
where the first term in (\ref{gamma}) is originated from quantum
vector multiplet $v^{++}_I$
\begin{equation}\label{gamma1}
\Gamma^{(1)}_v[V^{++}, q^+]=\frac{i}{2}\mbox{Tr}\ln(
\stackrel{\frown}{\Box}_{IJ} +q^{+a }\{T_I,T_J\}q^-_{a})~.
\end{equation}
Second term in (\ref{gamma}) is the contribution of ghosts and
quantum hypermultiplet $\xi^{+}_a$ and does not depend on the
background hypermultiplet.

As a result, the background hypermultiplet dependence of one-loop
effective action is included into the operator (\ref{box}), acting
on $v^{++}_I$ and containing the mass matrix of the vector
multiplet
\begin{equation}\label{mass1}
({\cal M}^{2}_{v})_{IJ} = \mbox{tr}\left([T_I,{\cal W}][\bar{\cal
W},T_J]+(I\leftrightarrow J)\right) + q^{+a}\{T_I,T_J\}q^{-}_{a}~,
\end{equation}
if $q^+$ is in the fundamental representation, and
\begin{equation}\label{mass2}
({\cal M}^{2}_{v})_{IJ} = \mbox{tr}\left([T_I,{\cal W}][\bar{\cal
W},T_J] + [q^{+a},T_I][T_J,q^{-}_{a}]\right) +(I\leftrightarrow
J)~,
\end{equation}
if $q^+$ in an arbitrary matrix representation.

In the above discussion, the gauge group structure of the
superfields ${\cal W}, q^+_a$ has been completely arbitrary.
Henceforth, the background superfields will be chosen to be
aligned along a fixed direction in the moduli space of vacua in
such a way that their scalar fields should solve Egs.
(\ref{vacua}). Then the hypermultiplet dependent effective action
in the case under consideration  takes the universal form
\begin{equation}\label{gamma2}
\Gamma^{(1)}_v[V^{++}, q^+]=\end{equation}$$\frac{i}{2}n(\Upsilon)
\times \mbox{Tr}\ln\left( \Box + \frac{i}{2}\alpha({H})({\cal
D}^+{\cal W}{\cal D}^- +\bar{\cal D}^+\bar{\cal W}\bar{\cal D}^- )
+ \alpha^2({H}) {\cal W}\bar{\cal W} + r(\Upsilon)q^{+a}q^{-}_{a}
\right)~.
$$
As the examples we list the values of $\alpha({H}), r(\Upsilon)$
and $n(\Upsilon)$ for models considered in \cite{KMT}.

$({\bf i})$ ${\cal N}=4$ SYM theory with gauge groups $\mathop{\rm
SU}(N)$, $\mathop{\rm Sp}(2N)$ and $\mathop{\rm SO}(N)$. Here the
hypermultiplet sector is composed of a single hypermultiplet in
the adjoint representation of the gauge group. The background was
chosen such that the gauge groups are broken down as follows
$\mathop{\rm SU}(N)\rightarrow \mathop{\rm SU}(N-1)\times
\mathop{\rm U}(1)$, $\mathop{\rm Sp}(2N)\rightarrow \mathop{\rm
Sp}(2N-2)\times \mathop{\rm U}(1)$, $\mathop{\rm SO}(N)\rightarrow
\mathop{\rm SO}(N-2)\times \mathop{\rm U}(1)$. All background
fields aligned along element $H=\mathop{\rm U}(1)$ of the Cartan
subalgebra (with $\Upsilon=H$). The mass matrix becomes
\begin{equation}
({\cal M}^2_v)_{IJ}=( {\cal W}\bar{\cal W}+ {\bf q}^{+a}{\bf
q}^-_a)(\alpha(H))^2 \delta_{I,J} \end{equation}
and traces in
Eq.(\ref{gamma}) produce the coefficient $n(\Upsilon)$ which is
equal to the number of roots with $\alpha(H)\neq 0$, i.e. to the
number of broken generators
$$
n(\Upsilon)=\cases{ 2(N-1)&\hbox {for} $\mathop{\rm SU}(N)$\,,\cr
4N-2&\hbox{for} $\mathop{\rm Sp}(2N)~{\rm and}~\mathop{\rm
SO}(2N+1)$\,,\cr 4N-1&\hbox{ for} $\mathop{\rm SO}(2N)$\,.\cr}
$$
The form of the mass matrix shows that in this case $r(\Upsilon) =
\alpha(H)$ .

$({\bf ii})$ The model introduced in \cite{Ah}. The gauge group is
$\mathop{\rm USp}(2N)=\mathop{\rm Sp}(2N,
{C})\bigcap \mathop{\rm U}(2N)$. The model contains four
hypermultiplets $q^+_F$ in the fundamental and one hypermultiplet
$q^+_A$ in the antisymmetric traceless representation $\mathop{\rm
USp}(2N)$. The background fields ${\cal W}$, $q^+_F$, $q^+_A$ are
chosen to solve Eqs. (\ref{vacua}) with the unbroken maximal gauge
subgroup $\mathop{\rm USp}(2N-2) \times \mathop{\rm U}(1)$:
$$
{\cal W}=\frac{{\bf \cal
W}}{\sqrt{2}}\,\mbox{diag}(1,\underbrace{0,...,0}_{N-1},
-1,\underbrace{0,...,0}_{N-1}), \quad q^+_F=0\,,
$$
$$
(q^+_A)_\alpha^{\;\;\beta}=\frac{\bf
q^+}{\sqrt{2N(N-1)}}\,\mbox{diag}(N-1,\underbrace{-1,...,-1}_{N-1},
N-1,\underbrace{-1,...,-1}_{N-1})\,.
$$
The mass matrix $({\cal M}^2_v)_{IJ}$ has been calculated in
\cite{KMT} and it has $n(\Upsilon)=4(N-1)$ eigenvectors with the
eigenvalue \begin{equation}{\cal M}^2_v=\bar{\bf \cal W}{\bf \cal
W}+\frac{N}{N-1}\bar{\bf q}^j{\bf q}_j.\end{equation}

$({\bf iii})$ The ${\cal N}=2$ superconformal model which is the
simplest quiver gauge theory \cite{Kach}. Gauge group is
$\mathop{\rm SU}(N)_L\times \mathop{\rm SU}(N)_R$. The model
contains two hypermultiplets $q^+$, $\tilde{q}^+$ in the
bifundamental representations $({ N},\bar{ N})$ and $(\bar{ N}, {
N})$ of the gauge group. In \cite{KMT} a solutions of
(\ref{vacua}) with non-vanishing hypermultiplet components that
specifies the flat directions in massless ${\cal N}=2$ SYM
theories has been constructed. The moduli space of vacua for this
model includes the following field configuration
$$
{\cal W}_L={\cal W}_R=\frac{ {\cal
W}}{N\sqrt{2(N-1)}}\,\mbox{diag}(N-1,
\underbrace{-1...,-1}_{N-1})\,,
$$
$$
q^+=\tilde{q}^+=\frac{{\bf
q}^+}{\sqrt{2}}\,\mbox{diag}(1,0,...,0)\,,
$$
which preserves an unbroken gauge group $\mathop{\rm
SU}(N-1)\times \mathop{\rm SU}(N-1)$ together with the diagonal
$\mathop{\rm U}(1)$ subgroup in $\mathop{\rm SU}(N)_L\times
\mathop{\rm SU}(N)_R$ associated with the chosen ${\cal W}$. In
such a background the mass matrix  has eigenvalue
\begin{equation}{\cal M}^2_v=\frac{1}{N-1}\bar{\cal W}{\cal W}
+\frac{1}{N}{\bf q}^{+a}{\bf q}^-_a\end{equation} and the
corresponding $n(\Upsilon)=4(N-1)$.

As the result, the hypermultiplet dependent effective action is
given by the expression (\ref{gamma2}). In the next section we
will consider the evaluation of this expression.

\section{Calculation of the one-loop effective action}
The expression (\ref{gamma2}) is a basis for an analysis of the
hypermultiplet dependence of the effective action. In the
framework of the  Fock - Schwinger proper-time representation, the
effective action (\ref{gamma2}) is written as follows
\begin{equation}\label{proper}
\Gamma^{(1)}_v [V^{++}, q^+]=\frac{i}{2} n(\Upsilon) \int d
\zeta^{(-4)} du \int^\infty_0 \frac{ds}{s}e^{-s(\Box
+\frac{i}{2}\alpha({H})({\cal D}^+{\cal W}{\cal D}^- +\bar{\cal
D}^+\bar{\cal W}\bar{\cal D}^-) + {\cal M}^2_v)}\times
\end{equation}
$$
\times ({\cal D}^+)^4
\left(\delta^{12}(z-z')\delta^{(-2,2)}(u,u')\right)|_{z=z',u=u'}=\int_0^\infty
\frac{d s}{s} \mbox{Tr} K(s),
$$
where ${\cal M}^2_v=\alpha^2({H}){\cal W}\bar{\cal W}+ r(\Upsilon)
q^{+a}q^-_a$. Here $K(s)$ is a superfield heat kernel, the
operation $\mbox{Tr}$ means the functional trace in the analytic
subspace of the harmonic superspace $\mbox{Tr}K(s)=\mbox{tr}\int
d\zeta^{(-4)}K(\zeta,\zeta|s)$, where $\mbox{tr}$ denotes the
trace over the discrete indices.  Representation of the effective
action (\ref{proper}) allows us to develop a straightforward
evaluation of the effective action in a form of covariant spinor
derivatives expansion in the superfield Abelian strengths ${\cal
W}, \bar{\cal W}$. The leading low-energy terms in this expansion
correspond to the constant space-time background $D^-_\alpha
D^+_\beta {\cal W}=\mbox{const}$, $\bar{D}^-_{\dot\alpha}
\bar{D}^+_{\dot\beta} \bar{\cal W}=\mbox{const}$ and on-shell
background hypermultiplet. However, it does not mean that we miss
all space-time derivatives in the component effective Lagrangian.
Grassmann measure in the integral over harmonic superspace
$d^4\theta^{+}d^4\theta^{-}$ generates four space-time derivatives
in component expansion of the superfield Lagrangian. Therefore the
above assumption is sufficient to obtain  a component effective
Lagrangian including four space-time derivatives of the scalar
components of the hypermultiplet.

Calculation of the effective action (\ref{proper}) is based on
evaluating the superfield heat kernel $K(s)$ and lead to a final
result for the hypermultiplet dependent low-energy one-loop
effective action of the Heisenberg-Euler type. We remind that the
whole background hypermultiplet is concentrated in ${\cal M}_v^2$.
The explicit form of it is:
\begin{equation}\label{manifest}
\Gamma^{(1)}[V^{++}, q^+]=\frac{1}{(4\pi)^2}n(\Upsilon)\int
d\zeta^{(-4)}du\int_0^\infty \frac{ds}{s^3}e^{-s(\alpha^2(H){\cal
W}\bar{\cal W}+r(\Upsilon)q^{+a}q^-_a)} \times
\end{equation}
$$
\times \frac{\alpha^4(H)}{16}(D^+{\cal W})^2(\bar{D}^+\bar{\cal
W})^2\frac{s^2({\cal N}^2-\bar{\cal N}^2)}{\cosh(s{\cal
N})-\cosh(s\bar{\cal N})}\cdot\frac{\cosh(s{\cal N})-1}{{\cal
N}^2}\cdot\frac{\cosh(s\bar{\cal N})-1}{\bar{\cal N}^2}~.
$$
Here ${\cal N}$ is given by ${\cal N}=\sqrt{-\frac{1}{2}D^4{\cal
W}^2}$. It can be expressed in terms of the two invariants of the
Abelian vector field ${\cal F}=\frac{1}{4}F^{mn}F_{mn}$ and ${\cal
G}=\frac{1}{4}{}^\star F^{mn}F_{mn}$ as ${\cal N}=\sqrt{2({\cal
F}+i{\cal G})}$.It is easily to see that the integrand in
(\ref{manifest}) can be expanded in power series in the quantities
$s^2{\cal N}^2$, $s^2\bar{\cal N}^2$. After change of proper time
$s$ to $s' {\cal W}\bar{\cal W}$ we get the expansion in power of
$s^{' 2}\frac{{\cal N}^2}{({\cal W}\bar{\cal W})^2}$ and their
conjugate. Since the integrand of (\ref{manifest}) is already
$\sim (D^+{\cal W})^2(\bar{D}^+\bar{\cal W})^2$, we can change in
each term of expansion the quantities ${\cal N}^2$, $\bar{\cal
N}^2$ by superconformal invariants $\Psi^2$ and $\bar\Psi^2$
\cite{BKT} expressing these quantities from $
\bar\Psi^2=\frac{1}{\bar{\cal W}^2}D^4\ln {\cal
W}=\frac{1}{2\bar{\cal W}^2}\{\frac{{\cal N}_\alpha^\beta{\cal
N}_\beta^\alpha}{{\cal W}^2}+ {\cal O}(D^+{\cal W})\} $ and its
conjugate. After that, one can show that each term of the
expansion can be rewritten as an integral over the full ${\cal
N}=2$ superspace.

It is interesting and instructive to evaluate the leading part of
the effective action (\ref{manifest}) that exactly coincides, up
to group factor $\Upsilon$ with the earlier results \cite{BI},
\cite{BP}, \cite{BIP}:
\begin{equation}\label{lead}
\Gamma^{(1)}_{\rm lead}=\frac{1}{(4\pi)^2}n(\Upsilon)\int
d^{12}z\, (\ln {\cal W}\ln \bar{\cal W}+
\mbox{Li}_2(X)+\ln(1-X)-\frac{1}{X}\ln(1-X)).
\end{equation}
Here $\mbox{Li}_{2}(X)$ is the Euler's dilogarithm function.
Next-to-leading corrections to (\ref{lead}) can also be
calculated. The remarkable feature of the low-energy effective
action (\ref{lead}) is the appearance of the factor
$r(\Upsilon)/\alpha({H})$ in argument $X$. This factor is
conditioned by the vacuum structure of the model under
consideration and depends on the specific features of the symmetry
breaking.

Now we discuss some terms in the component Lagrangian
corresponding to the effective action (\ref{lead}). Component
structure of the effective action (\ref{lead}) has been studied
\cite{BI}   in the context of ${\cal N}=4$ SYM theory in bosonic
sector for completely constant background fields $F_{mn}, \phi,
\bar\phi, f^i, \bar{f}_i$. However, it was pointed out above that
the superfield effective action (\ref{lead}) allows us to find the
terms in the effective action up to fourth order in space-time
derivatives of component fields. Now our aim is to find such terms
in the hypermultiplet scalar component sector. To do that we omit
all components of the background superfields besides the scalars
$\phi, {\bar{\phi}}$ in the ${\cal N}=2$ vector multiplet and
scalars $f, \bar{f}$ in the hypermultiplet and integrate over
$d^4\theta^{+}d^4\theta^{-}=(D^-)^4(D^+)^4$.  To get the leading
space-time derivatives of the hypermultiplet scalar components we
should put exactly two spinor derivatives on each hypermultiplet
superfield. It yields, after some transformations, to the
following term with four space-time derivatives on $q^{\pm}$ in
component expansion of effective action :
$$
\Gamma^{(1)}_{\rm lead}=\int
d^4xdu\frac{n(\Upsilon)}{(4\pi)^2}\sum_{k=2}^\infty\frac{1}{16}\frac{k-1}{k(k+1)}
\frac{X^{k-2}}{({\cal W}\bar{\cal W})^2}\times$$
$$\{-\bar{D}^{+\dot\alpha}D^{+\alpha}q^-_b
\bar{D}^+_{\dot\alpha}D^-_\beta
q^{+(b}\bar{D}^{-\dot\beta}D^{-\beta}q^{+a)}\bar{D}^-_{\dot\beta}D^+_\alpha
q^-_a
$$
$$
+\frac{1}{2}\bar{D}^{+\dot\alpha}D^{+\alpha}q^-_b
\bar{D}^{-\dot\beta}D^{-\beta}
q^{+b}\bar{D}^-_{\dot\beta}D^-_\beta
q^{+a}\bar{D}^+_{\dot\alpha}D^+_\alpha q^-_a$$$$
+\frac{1}{2}\bar{D}^{-\dot\beta}D^{+\alpha}q^-_b
\bar{D}^{+\dot\alpha}D^{-\beta}
q^{+b}\bar{D}^+_{\dot\alpha}D^-_\beta
q^{+a}\bar{D}^-_{\dot\beta}D^+_\alpha q^-_a\}|_{\theta=0}\,.
$$
The straightforward calculation of the components in this
expression shows that among the many terms with four derivatives
there is an interesting term of the special type. As the first
term in expansion over variable
$X_0=\frac{r(\Upsilon)\bar{f}^if_i}{\alpha^2\bar\phi\phi}$ we have
\begin{equation}\label{chern}
\Gamma^{(1)}_{\rm lead}=-\frac{1}{48\pi^2}
n(\Upsilon)\left(\frac{r(\Upsilon)}{\alpha({H})}\right)^2
\int d^4x \end{equation}$$\times
\frac{1}{(\phi\bar\phi)^2}i\varepsilon^{\mu\nu\lambda\rho}(\partial_\mu
\bar{f}^i\partial_\nu f_i \partial_\lambda \bar{f}^j \partial_\rho
f_j -\partial_\mu \bar{f}^i\partial_\nu \bar{f}_i \partial_\lambda
{f}^j \partial_\rho f_j)$$

 The expression
(\ref{chern}) has a form of the Chern-Simons-like action for the
multicomponent complex scalar filed. The terms of such form in the
effective action were discussed in Refs. \cite{TZ}, \cite{arg} in
context of ${\cal N}=4,2$ SYM models and in Refs. \cite{Int} for
$d=6, {\cal N}=(2,0)$ superconformal models respectively. Here the
expression (\ref{chern}) is obtained as a result of
straightforward calculation in the supersymmetric quantum field
theory.

\section{Hypermultiplet dependent contribution \\to
the effective action beyond the on-shell condition}

In the above consideration a crucial point was the condition that
the hypermultiplet $q^+$ satisfies the one-shell conditions
(\ref{onsh}) and the constraint $q^+=D^{++}q^-$. Here we relax the
on-shell conditions and study some of possible subleading
contributions with the minimal number of space-time derivatives in
the component effective action.

\begin{figure} [ht]
\begin{center}
\includegraphics{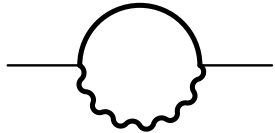}
\caption{One-loop supergraph} \label{f1}
\end{center}
\end{figure}

We consider a supergraph given in Fig.\ref{f1} with two external
hypermultiplet legs and with all propagators depending on the
background ${\cal N}=2$ vector multiplet. Here the wavy line
stands for the ${\cal N}=2$ gauge superfield propagator and the
solid external and internal lines stand for the background
hypermultiplet superfields and quantum hypermultiplet propagator
respectively. For simplicity we suppose that the background field
is Abelian and omit all group factors. The corresponding
contribution to effective action looks like
\begin{equation}\label{gamma_2}
i\Gamma_2 =\int d \zeta_1^{(-4)} d \zeta_2^{(-4)} du_1
du_2\left(\frac{({\cal D}^+_1)^4 ({\cal D}^+_2)^4}{(u^+_1
u^+_2)^3}\frac{1}{{\stackrel{\frown}{\Box}}_1}\delta^{12}(1|2)\right)
\times
\end{equation}
$$
\times \left(\frac{({\cal D}^+_2)^4 ({\cal
D}^+_1)^4}{{\stackrel{\frown}{\Box}}_2{\stackrel{\frown}{\Box}}_1}
\delta^{12}(2|1)({\cal D}_1^{--})^2
\delta^{(-2,2)}(u_2,u_1)\right)
\tilde{q}^{+}(z_1,u_1)q^+(z_2,u_2).
$$
As usually, we extract the factor $(D^+)^4$ from the vector
multiplet propagator for reconstructing the full ${\cal N}=2$
measure. Then we shrink a loop into a point by transferring the
$\stackrel{\frown}{\Box}$ and $({\cal D}^+)^4$ from first
$\delta$-function to another one and kill one integration. At this
procedure the operator $\stackrel{\frown}{\Box}$ does not act on
$q^+$ because we are interesting in the minimal number of
space-time derivatives in the component form of the effective
action. As a result, one obtains
\begin{equation}
i\Gamma_2 =\left.\int \frac{d \zeta_1^{(-4)}  du_1 du_2}{(u^+_1
u^+_2)^3} \frac{({\cal D}^+_1)^4({\cal D}^+_2)^4 ({\cal
D}^+_1)^4}{{\stackrel{\frown}{\Box}}_2{\stackrel{\frown}{\Box}}_1^2}
\delta^{12}(z-z')\right|\times \end{equation}
$$
\times \left( ({\cal D}_1^{--})^2 \delta^{(-2,2)}(u_2,u_1)\right)
\tilde{q}^+(z_1,u_1)q^+(z_1,u_2)~.
$$
Further we use twice the relation (\ref{polin}) allowing us to
express the $({\cal D}^{+}_1)^4({\cal D}^+_2)^4$ as a polynomial
in powers of $(u^+_1u^+_2)$.  Then after multiplying the $({\cal
D}^+_1)^4({\cal D}^+_2)^4 ({\cal D}^+_1)^4$ with the distribution
$1/(u^+_1u^+_2)^3$ we obtain a polynomial in $(u^+_1u^+_2)$
containing the powers of this quantity from 5-th to 1-st. The
first order is just a contribution of the type which we considered
in the previous section, because one derivation $(D^{--})^2$ is
used for transformation $(u^+_1u^+_2)$ into
$(u^+_1u^-_2)|_{u_1=u_2} =1$ in the coincident limit. Another
$D^{--}$ transforms $q^+$ into $q^-$. All that has been already
done in Section 4.

Here we consider the new contribution to the effective action
containing term $(u^+_1u^+_2)^2$ in the above polynomial:
\begin{equation}
\frac{({\cal D}^+_1)^4({\cal D}^+_2)^4 ({\cal D}^+_1)^4}{(u^+_1
u^+_2)^3}=\end{equation}$$...+ (u^+_1
u^+_2)^2(u^-_1u^+_2)(u^-_2u^+_1)({\cal
D}^+_1)^4\left(\frac{i}{2}{\stackrel{\frown}{\Box}}_1
\Delta^{--}_2 (u^+_2u^-_1)-\frac{i}{2} \Delta^{--}_1
{\stackrel{\frown}{\Box}}_2 (u^+_1u^-_2)\right)+...
$$
The ellipsis means the terms with the powers of $(u^+_1u^+_2)$
other then 2. One can show that in the coincident limit they
disappear. Now transferring $(D^{--})^2$ on $(u^+_1u^+_2)^2$ we
obtain the expression:
\begin{equation}
i\Gamma_2 = i \int d\zeta^{(-4)}du ({\cal D}^+)^4
\frac{1}{{\stackrel{\frown}{\Box}}^3}(\underbrace{{\stackrel{\frown}{\Box}}\Delta^{--}}_{\Gamma_2(1)}-
\underbrace{\Delta^{--}{\stackrel{\frown}{\Box}}}_{\Gamma_2(2)})\delta^{12}(z-z')|_{z=z'}\tilde{q}^+(z,u)q^+(z,u)\,,
\end{equation}
where $\Delta^{--}$ is defined in (\ref{Delt}).

Let us consider each of the two underlined contributions
separately. We use the representation
\begin{equation}\label{repr}
\frac{1}{{\stackrel{\frown}{\Box}}^2}\Delta^{--}\delta^{12}(z-z')|=\int
ds\, s e^{s{\stackrel{\frown}{\Box}}}\Delta^{--}
\delta^{12}(z-z')|,
\end{equation}
where $|$ means the coincident limit $z=z'$. Then we  can apply a
derivative expansion of the heat kernel. The goal is to collect
the maximum possible number of factors of ${\cal D}^+, {\cal D}^-$
acting on $(\theta^+ -\theta^{'+})^4(\theta^- -\theta^{'-})^4$ and
having the minimum order in $s$ in the integral over $s$. Higher
orders in $s$ generate the higher spinor derivatives in the
effective action. We take terms $\frac{1}{2}{\cal W}({\cal D}^-)^2
+c.c.$ from $\Delta^{--}$ and expand the exponential so as to find
$({\cal D}^-)^4$. The Eq. (\ref{repr}) allows us to write the
leading contribution to $\Gamma_2(1)$ as follows
\begin{equation} \Gamma_2(1)=-\int d^{12}zdu\int^\infty_0ds \cdot s\int
\frac{d^4p}{(2\pi)^4}e^{-sp^2}e^{s({\cal W}\bar{\cal
W}-\varepsilon)}\frac{s^2}{32} \bar{\cal W}(D^{+\alpha}{\cal
W}D^+_\alpha{\cal W})\times \end{equation}$$\times
(D^-)^2(\bar{D}^-)^2\delta^8(\theta-\theta')| \tilde{q}^+ q^+ +
\hbox{ c.c.}
$$
After trivial integration over p and s this contribution has the
form
\begin{equation}\label{gamma(1)}
\Gamma_2(1)=\frac{i}{32\pi^2}\int d^{12}z du \frac{D^+{\cal
W}D^+{\cal W}}{\bar{\cal W}{\cal W}^2}
\tilde{q}^+(z,u)q^+(z,u)({\cal D}^-)^4 \delta^{8}(\theta-\theta')|
+c.c.
\end{equation}

Now we fulfil the same manipulations with the second underlined
contribution $\Gamma_2(2)$ keeping the same order in $s$ and $D^-,
\bar{D}^-$ as in the expression (\ref{gamma(1)}). After that we
see that the leading term of the form (\ref{gamma(1)}) is absent
in $\Gamma_2(2)$. Then it is not difficult to show that the
contribution (\ref{gamma(1)}) is rewritten as follows [we use
$\int d^2\bar\theta^-=\bar{D}^{+2}$]
$$
-\frac{i}{32\pi^2}\int d^4x d^4\theta^+ d^2\theta^- du
(\bar{D}^+)^2 (D^+)^2 \frac{\ln{\cal W}}{\bar{\cal
W}}\tilde{q}^+(z,u)q^+(z,u)({\cal D}^-)^4
\delta^{8}(\theta-\theta')|
$$
The non-zero result arises when all $D^{+}$ - factors act only on
the spinor delta-function. Thus, the contribution under
consideration is written as an integral over the measure
$d^4xdud^4{\theta}^+d^2{\theta}^-$ which looks like
 "$3/4$ - part" of the full ${\cal N}=2$ harmonic superspace measure
$d^4xdud^4{\theta}^+d^4{\theta}^-$.

Therefore, the hypermultiplet dependent effective action contains
the term
\begin{eqnarray}\label{3/4}
\Gamma_2=&-&\frac{i}{32\pi^2}\int d^4xdud^4\theta^+ d^2\theta^-
\frac{1}{\bar{\cal W}} \ln ({\cal W})\tilde{q}^+q^+ |_{\bar\theta^-=0}\\
&-&\frac{i}{32\pi^2}\int d^4xdud^4\,\theta^+ d^2\bar\theta^-
\frac{1}{{\cal W}} \ln (\bar{\cal
W})\tilde{q}^+q^+|_{\theta^-=0}~.\nonumber
\end{eqnarray}
Presence of such a term in the effective action for ${\cal N}=2$
supersymmetric models in subleading order was proposed in
\cite{arg}. Here we have shown how this term can be derived in the
supersymmetric quantum field theory.

It is interesting and instructive to find a component form of such
a non-standard superfield action (\ref{3/4}). Here we consider
only a purely bosonic sector of (\ref{3/4}). After integration
over anticommuting variables, which can be equivalently replaced
by supercovariant derivatives evaluated at $\theta=0$,  we obtain
a Chern-Simons-like contribution to the effective action
containing three space-time derivatives
\begin{equation}\label{last}
\Gamma_2= -\frac{1}{2\pi^2}\int d^4x
\frac{1}{\phi\bar\phi}\varepsilon^{mnab}\partial_m \bar{f}^i
\partial_n f_i F_{ab}~.
\end{equation}
This expression is the simplest contribution to the hypermultiplet
dependent effective action beyond the on-shell conditions
(\ref{onsh}) for the background hypermultiplet. Of course, there
exist other, more complicated contributions including the
hypermultiplet derivatives, they also can be calculated by the
same method which led to (\ref{3/4}). Here we only demonstrated a
procedure which allows us to derive the contributions to the
effective action in the form of integral over $3/4$ - part of the
full ${\cal N}=2$ harmonic superspace.

\section{Summary}
We have studied the one-loop low-energy effective action in ${\cal
N}=2$ superconformal models. The models are formulated in harmonic
superspace and their field content correspond to the finiteness
condition (\ref{fin}). Effective action depends on the background
Abelian ${\cal N}=2$ vector multiplet superfield and background
hypermultiplet superfields satisfying the special restrictions
(\ref{vacua}), (\ref{vac}) which define the vacuum structure of
the models. The effective action is calculated on the base of the
${\cal N}=2$ background field method for the background
hypermultiplet on-shell (\ref{onsh}) and beyond the on-shell
conditions. For an on-shell hypermultiplet we found the universal
expression for the effective active action. For hypermultiplet
beyond on-shell, we calculated the special manifestly ${\cal N}=2$
supersymmetric subleading contribution which is written as an
integral over $3/4$ of the full ${\cal N}=2$ harmonic superspace.
We believe that such contributions deserves a special study.

\section*{Acknowledgments}
N.G.P is grateful to I.L. Buchbinder for collaboration and S.
Kuzenko and I. McArthur for helpful discussions and
correspondence. The work was supported in part by RFBR grants,
project No 06-02-16346, No 08-02-00334-a, grant for LRSS, project
No 2553.2008.2 and INTAS grant, project  No 05-7928.

\end{document}